# Solid-state heteronuclear multiple-quantum spectroscopy under a magic-angle spinning frequency of 150 kHz

Eric Chung-Yueh Yuan,[a] Po-Wen Chen,[a] Shing-Jong Huang,[b] Mai-Liis Org,[c] Ago Samoson,[c,*] and Jerry Chun Chung Chan[a,*]

[a] Department of Chemistry, National Taiwan University, Taipei 10617, Taiwan, Republic of China

[b] Instrumentation Center, National Taiwan University, Taipei 10617, Taiwan, Republic of China

[c] Tallinn University of Technology, Estonia

**Abstract**

We hereby demonstrate that $^1$H detected $^{15}$N-$^1$H heteronuclear multiple-quantum spectroscopy can be carried out at a magic angle spinning frequency of 150 kHz. While the $^{15}$N-$^1$H multiple-quantum coherences can be directly excited from the dipolar order created by the method of adiabatic demagnetization in the rotating frame, it is technically more advantageous to acquire the chemical shift evolution of the heteronuclear multiple-quantum coherence by two separate chemical shift evolution periods for $^1$H and $^{15}$N. We also show that the heteronuclear multiple-quantum correlation spectrum can be obtained by shearing the corresponding heteronuclear single-quantum correlation spectrum.

## 1  INTRODUCTION

Solid-state $^1$H NMR spectroscopy under the condition of magic angle spinning (MAS) has been established as an important analytical method for a large variety of systems including pharmaceuticals, supramolecular structures, inorganic–organic hybrid materials, and biological solids.[1,2] It has been shown that indirect detection of solid-state $^{13}$C and $^{15}$N NMR could be achieved efficiently by $^1$H detection under MAS conditions.[3,4] While MAS is arguably the most advantageous strategy for the resolution improvement in $^1$H NMR,[5] the resolution of solid-state $^1$H spectra can be further improved by correlation spectroscopy.[6,7] Very recently, heteronuclear single-quantum correlation spectroscopy (HSQC) had been successfully applied for the study of protein aggregates and wood samples under MAS frequencies in the regime of 150 kHz.[8,9] In view of the considerable interest in heteronuclear multiple-quantum (MQ) spectroscopy,[10–14] we attempt to investigate the feasibility of carrying out correlation spectroscopy involving the heteronuclear MQ coherences under a spinning frequency of 150 kHz. As justified by theoretical consideration and illustrated by experimental results, we found that the MQ correlation spectra could be acquired with and without direct excitation of the MQ coherences.

---

* Corresponding authors: ago.samoson@ttu.ee, chanjcc@ntu.edu.tw

## 2 THEORY

### 2.1 Resolution enhancement on multiple-quantum dimension

Taking the spin system containing $^1$H and $^{15}$N for an instance, the chemical shift of a heteronuclear MQ coherence is defined as:

$$\omega_{MQ} = \omega_N + p\omega_H \tag{1}$$

where $p$ equal to +1 and −1 correspond to the cases of double-quantum (DQ) and zero-quantum (ZQ) coherences, respectively. The resolution factor of a signal pair ($R_{12}$) is defined as:

$$R_{12} = \frac{|\omega_1 - \omega_2|}{\sqrt{\Delta_1^2 + \Delta_2^2}} \tag{2}$$

where $\omega_k$ and $\Delta_k$ are the chemical shift and line width at half maximum of signal $k$, respectively. As shown by two generic spectra corresponding to the $^{15}$N spectrum and the $^{15}$N-$^1$H DQ spectrum in Figure 1, the resolution of the DQ spectrum is superior to the $^{15}$N spectrum. Although it is not always the case that a heteronuclear DQ spectrum has better resolution than the SQ analog, it is of interest to investigate how to acquire heteronuclear MQ (hetMQ) spectra. There are many methods to excite MQ coherences. Under the conditions of MAS, one can excite MQ coherences by adiabatic passage.[15,16] We will first illustrate the idea of adiabatic passage in the context of cross-polarization (CP), based on the formalism developed by Nielsen and co-workers.[17,18] After that, we will discuss how the method can be exploited to excite MQ coherences.

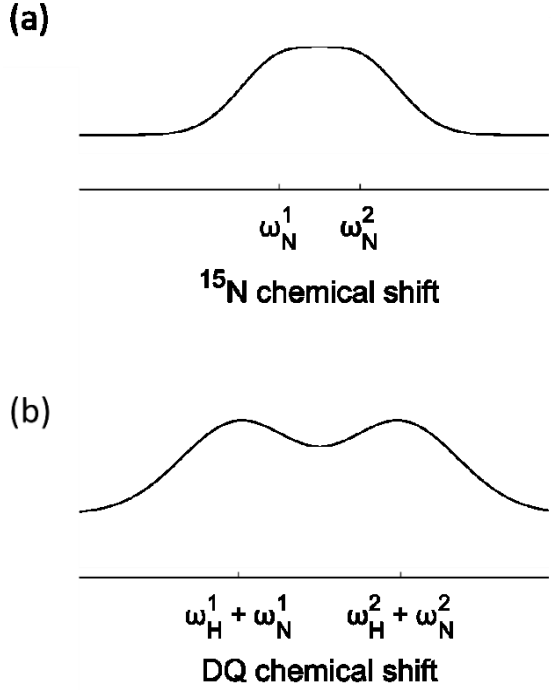

**Figure 1.** Schematic (a) $^{15}$N and (b) $^{15}$N-$^{1}$H double-quantum spectra for a spin system containing two $^{1}$H-$^{15}$N spin pairs with $^{1}$H chemical shifts of 100 and −100 Hz and $^{15}$N chemical shifts of 100 and −100 Hz, respectively. A line broadening of 100 Hz was applied for both spectra.

## 2.2 Adiabatic passage through the Hartmann−Hahn condition

Consider an isolated heteronuclear spin-1/2 pair $I$ and $S$. Under MAS conditions, the Hamiltonian in the double rotating frame is

$$H(t) = \underbrace{\omega_I I_z + \omega_S S_z}_{H_{CS}} + \underbrace{\omega_{IS}(t) 2 I_z S_z}_{H_D} + \underbrace{\omega_{1I}(t) I_x + \omega_{1S}(t) S_x}_{H_{RF}} \tag{3}$$

where $\omega_I$, $\omega_S$, and $\omega_{IS}(t)$ represent the isotropic chemical shifts (CS) of spins $I$ and $S$, and their time-dependent dipolar coupling constant, respectively; $\omega_{1I}(t)$ and $\omega_{1S}(t)$ denote the amplitude-modulated radio-frequency (RF) fields applied on the $I$ and $S$ channels. At a spinning frequency of $\omega_r$, $\omega_{IS}(t)$ can be expanded in a Fourier series as follows:

$$\omega_{IS}(t) = \sum_{m=-2}^{2} \omega_m e^{im\omega_r t}$$

$$\begin{cases} \omega_{\pm 1} = -\dfrac{d_{IS}}{2\sqrt{2}} \sin 2\beta \, e^{\pm i\gamma} \equiv \kappa_1 e^{\pm i\gamma} \\ \omega_{\pm 2} = \dfrac{d_{IS}}{4} \sin^2 \beta \, e^{\pm 2i\gamma} \equiv \kappa_2 e^{\pm 2i\gamma} \end{cases} \tag{4}$$

where $m \neq 0$, the dipolar coupling constant $d_{IS}$ is $-\mu_0 \gamma_I \gamma_S / 4\pi r_{IS}^3$, and the Euler angles $\beta$ and $\gamma$ describe the direction of the internuclear vector $\vec{r}_{IS}$ with respect to the spinning axis.

To describe the Hamiltonian of the spin system under MAS conditions, we transform the system into the interaction frame defined by two constant RF fields, viz., $\omega_{1I}^0$ and $\omega_{1S}^0$:[17,18]

$$\tilde{H}(t) = U_{RF}^{-1}(t)H(t)U_{RF}(t) = \tilde{H}_{CS}(t) + \tilde{H}_D(t) + \tilde{H}_{RF}(t) \tag{5}$$

where

$$U_{RF}(t) = e^{-i(\omega_{1I}^0 t I_x + \omega_{1S}^0 t S_x)} \tag{6}$$

In particular, the chemical shift term, viz., $\tilde{H}_{CS}(t)$, is explicitly written as

$$\tilde{H}_{CS}(t) = U_{RF}^{-1}(t)H_{CS}(t)U_{RF}(t)$$

$$= \frac{1}{2}\left[\omega_I\left(I_+ e^{-i\omega_{1I}^0 t} + I_- e^{i\omega_{1I}^0 t}\right) + \omega_S\left(S_+ e^{-i\omega_{1S}^0 t} + S_- e^{i\omega_{1S}^0 t}\right)\right] \tag{7}$$

where $I_\pm = I_z \pm iI_y$ and $S_\pm = S_z \pm iS_y$. Its time average becomes zero under the first-order approximation, i.e., $\overline{\tilde{H}}_{CS}^{(1)} = 0$, because the isotropic chemical shifts ($\omega_I$, $\omega_S$) are time independent. On the other hand, the second term of Equation (5), viz., the dipolar term is calculated as

$$\tilde{H}_D(t) = U_{RF}^{-1}(t)H_D(t)U_{RF}(t)$$

$$= \sum_{m=-2}^{2} \frac{\omega_m}{2} e^{im\omega_r t}\left[I_+ S_+ e^{-i(\omega_{1I}^0 + \omega_{1S}^0)t} + I_- S_- e^{i(\omega_{1I}^0 + \omega_{1S}^0)t}\right.$$

$$\left. + I_+ S_- e^{-i(\omega_{1I}^0 - \omega_{1S}^0)t} + I_- S_+ e^{i(\omega_{1I}^0 - \omega_{1S}^0)t}\right] \tag{8}$$

It can have non-vanishing time-averaged values, provided that the time dependence of $\omega_{IS}(t)$ is matched to the frequencies of the oscillation terms in the bracket. For the particular condition of $\omega_{1I}^0 - \omega_{1S}^0 = \omega_r$, its first-order time average is calculated as

$$\overline{\tilde{H}}_D^{(1)} = \frac{1}{2}(\omega_{+1}I_+ S_- + \omega_{-1}I_- S_+) \tag{9}$$

Other non-vanishing time averages can be obtained for $\omega_{1I}^0 - \omega_{1S}^0 = 2\omega_r$ and so on. The conditions of $\omega_{1I}^0 - \omega_{1S}^0 = m\omega_r$, $m = \pm1, \pm2$ are known as the first-order Hartmann–Hahn sideband matching conditions, or for brevity, the ZQ matching condition comprising the $I_+ S_-$ and $I_- S_+$ operators. Accordingly, the conditions of $\omega_{1I}^0 + \omega_{1S}^0 = m\omega_r$, $m = \pm1, \pm2$ are referred to as the matching conditions that recouple the DQ Hamiltonian or the DQ matching condition containing the operators of $I_+S_+$ and $I_-S_-$.

It proves useful to construct the ZQ and DQ subspaces by introducing the fictitious spin-1/2 operators (*vide infra*).[19,20] Following the notation of Khaneja et al.,[21] the DQ subspace is spanned by the following operators:

$$I_x^{DQ} = \frac{1}{2}(I_x + S_x)$$
$$I_z^{DQ} = \frac{1}{2}(I_+S_+ + I_-S_-) = I_zS_z - I_yS_y$$
$$I_y^{DQ} = \frac{1}{2i}(I_+S_+ - I_-S_-) = I_yS_z + I_zS_y \tag{10}$$

whereas the ZQ subspace is spanned by

$$I_x^{ZQ} = \frac{1}{2}(I_x - S_x)$$
$$I_z^{ZQ} = \frac{1}{2}(I_+S_- + I_-S_+) = I_zS_z + I_yS_y$$
$$I_y^{ZQ} = \frac{1}{2i}(I_+S_- - I_-S_+) = I_yS_z - I_zS_y \tag{11}$$

Accordingly, Equation (8) can be rewritten as

$$\tilde{H}_D(t) = \sum_{m=-2}^{2} \kappa_m \times$$
$$\{I_z^{DQ} \cos[m\gamma + m\omega_r t - (\omega_{1I}^0 + \omega_{1S}^0)t]$$
$$+ I_y^{DQ} \sin[m\gamma + m\omega_r t - (\omega_{1I}^0 + \omega_{1S}^0)t]$$
$$+ I_z^{ZQ} \cos[m\gamma + m\omega_r t - (\omega_{1I}^0 - \omega_{1S}^0)t]$$
$$+ I_y^{ZQ} \sin[m\gamma + m\omega_r t - (\omega_{1I}^0 - \omega_{1S}^0)t]\} \tag{12}$$

where the scaling factor $\kappa_m$ denotes the amplitude of $\omega_m$, as indicated in Equation (4). Similarly, Equation (9) is rewritten as

$$\overline{\tilde{H}}_D^{(1)} = \kappa_1(I_z^{ZQ} \cos\gamma + I_y^{ZQ} \sin\gamma) \equiv \kappa_1 I_D^{ZQ} \tag{13}$$

where the operator $I_D^{ZQ}$ lies on the *yz* plane of the ZQ subspace and is always perpendicular to $I_x^{ZQ}$. The amplitude of this Hamiltonian is independent of the *γ* angle.[22] This feature of the average Hamiltonian can be achieved by the class of γ-encoded recoupling sequences.[17,18] Because $I_D^{ZQ}$ is orthogonal to $I_x$, $I_y$, and $I_z$, the density operator proportional to $I_D^{ZQ}$, which must commute with the Hamiltonian defined in Equation (13), is commonly referred to as the dipolar-order state.[23,24]

We now turn our attention to the third term of Equation (5), viz., $\tilde{H}_{RF}(t)$, which is calculated as

$$\widetilde{H}_{RF}(t) = U_{RF}^{-1}(t)H_{RF}(t)U_{RF}(t) - (\omega_{1I}^0 t I_x + \omega_{1S}^0 t S_x)$$

$$= [\omega_{1I}(t) - \omega_{1I}^0]I_x + [\omega_{1S}(t) - \omega_{1S}^0]S_x \qquad (14)$$

Again, we can rewrite the Hamiltonian in terms of the basis operators of the ZQ and DQ subspaces:

$$\widetilde{H}_{RF}(t) = [\omega_{1I}(t) + \omega_{1S}(t) - \omega_{1I}^0 - \omega_{1S}^0]I_x^{DQ} + [\omega_{1I}(t) - \omega_{1S}(t) - \omega_{1I}^0 + \omega_{1S}^0]I_x^{ZQ} \qquad (15)$$

Under the ZQ matching condition of $\omega_{1I}^0 - \omega_{1S}^0 = \omega_r$, we have

$$\widetilde{H}_{RF}(t) = [\omega_{1I}(t) + \omega_{1S}(t) - \omega_{1I}^0 - \omega_{1S}^0]I_x^{DQ} + [\omega_{1I}(t) - \omega_{1S}(t) - \omega_r]I_x^{ZQ} \qquad (16)$$

We now have all the required expressions to construct a useful approximation for the Hamiltonian in the interaction frame under the ZQ matching condition. Because the RF term is varied at a significantly slower timescale than the chemical shift and dipolar terms, it is legitimate to approximate the latter two terms by their time averages. As shown in Equation (7), the time average of the chemical shift term vanishes, whereas the time-averaged dipolar term is given by Equation (13). Thus, Equation (5) can be approximated as:

$$\widetilde{H}(t) \approx \overline{\widetilde{H}}_D^{(1)} + \widetilde{H}_{RF} = \widetilde{H}^{DQ}(t) + \widetilde{H}^{ZQ}(t) \qquad (17)$$

where

$$\widetilde{H}^{DQ}(t) = [\omega_{1I}(t) + \omega_{1S}(t) - \omega_{1I}^0 - \omega_{1S}^0]I_x^{DQ} \qquad (18)$$

and

$$\widetilde{H}^{ZQ}(t) = [\omega_{1I}(t) - \omega_{1S}(t) - \omega_r]I_x^{ZQ} + \kappa_1 I_D^{ZQ} \qquad (19)$$

For the ZQ part, we can define an effective field $I_e^{ZQ}(t)$ as follows:

$$\widetilde{H}^{ZQ}(t) = \omega_e(t)I_e^{ZQ}(t)$$

$$I_e^{ZQ}(t) = I_x^{ZQ}\cos\theta(t) + I_D^{ZQ}\sin\theta(t)$$

$$\omega_e(t) \equiv \sqrt{[\omega_{1S}(t) - \omega_{1I}(t) - \omega_r]^2 + \kappa_1^2}$$

$$\theta(t) \equiv \tan^{-1}\left[\frac{\kappa_1}{\omega_{1I}(t) - \omega_{1S}(t) - \omega_r}\right] \qquad (20)$$

The inclination angle $\theta(t)$ can be varied continuously by $\omega_{1I}(t)$ and $\omega_{1S}(t)$. With reference to the ZQ matching condition, three limiting cases of the inclination angle $\theta(t)$ can be obtained. For $\omega_{1I} - \omega_{1S} \gg \omega_r$, we have $\theta = 0$. Thus,

$$\widetilde{H}^{ZQ}(t) = [\omega_{1I}(t) - \omega_{1S}(t) - \omega_r]I_x^{ZQ} \qquad (21)$$

For $\omega_{1I} - \omega_{1S} = \omega_r$ (ZQ Hartmann-Hahn sideband matching), we have $\theta = \pi/2$. Hence,

$$\tilde{H}^{ZQ}(t) = \kappa_1 I_D^{ZQ} \tag{22}$$

For $\omega_{1I} - \omega_{1S} \ll \omega_r$, we have $\theta = \pi$ and

$$\tilde{H}^{ZQ}(t) = -|\omega_{1I}(t) - \omega_{1S}(t) - \omega_r| I_x^{ZQ} \tag{23}$$

Assume the initial state of the spin system ($\rho_i$) after a π/2 excitation is $I_x$, which can be rewritten in the interaction frame as

$$\tilde{\rho}_i = U_{RF}^{-1}(t)\rho_i U_{RF}(t) = I_x^{DQ} + I_x^{ZQ} \tag{24}$$

Under the influence of the $\tilde{H}(t)$ in Equation (17), it can be shown that $I_x^{DQ}$ remains stationary, whereas $I_x^{ZQ}$ would evolve in the ZQ subspace. In particular, we consider the evolution pathway of $\theta = 0 \to \theta = \pi/2 \to \theta = \pi$. In other words, the RF field mismatch ($\omega_{1I} - \omega_{1S}$) is initially much larger than the spinning frequency ($\omega_r$). The mismatch is then gradually reduced so that it slowly approaches the matching condition ($\theta = \pi/2$). If the process is slow enough (adiabatic), the spin system will evolve to $\tilde{\rho}_D$ as follows:

$$\tilde{\rho}_i = I_x^{DQ} + I_x^{ZQ} \xrightarrow{\theta=0 \to \frac{\pi}{2}} \tilde{\rho}_D = I_x^{DQ} + I_D^{ZQ} \tag{25}$$

This process is commonly referred to as adiabatic demagnetization in the rotating frame (ADRF). That is, the ADRF process can generate a dipolar state ($I_D^{ZQ}$), which does not have any components of $I_x$, $I_y$, nor $I_z$ (Zeeman order). Next, the RF field mismatch is slowly adjusted to reach a large negative value, i.e., $\omega_{1I} - \omega_{1S} \ll \omega_r$ ($\theta = \pi$). If the change in the Hamiltonian is again adiabatic, $\tilde{\rho}_D$ will evolve further to the final state ($\tilde{\rho}_f$):

$$\tilde{\rho}_D = I_x^{DQ} + I_D^{ZQ} \xrightarrow{\theta=\frac{\pi}{2} \to \pi} \tilde{\rho}_f = I_x^{DQ} - I_x^{ZQ} = S_x \tag{26}$$

The process is known as adiabatic remagnetization in the rotating frame (ARRF). Both the processes of ADRF and ADRF are reversible. In the double rotating frame, that is

$$\rho_f = U_{RF}(t)\tilde{\rho}_f U_{RF}^{-1}(t) = S_x \tag{27}$$

because $\tilde{\rho}_f$ commutes with $U_{RF}(t)$. As a result, the sequential process comprising ADRF and ARRF, which is referred to as adiabatic passage through the Hartmann–Hahn condition (APHH), can be exploited to achieve the polarization transfer from $I_x$ to $S_x$ and vice versa, via the dipolar state $I_D^{ZQ}$. The trajectory is pictorially depicted in Figure 2.

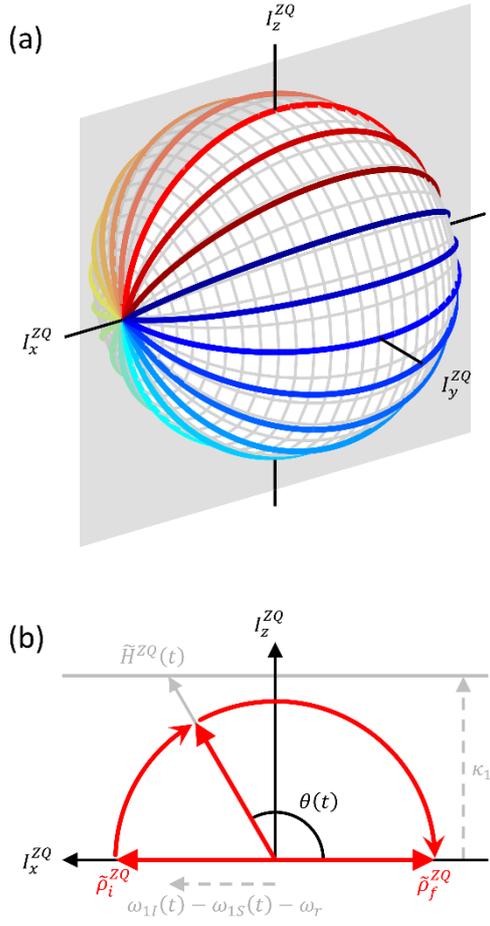

**Figure 2.** Trajectories of the state (density operator) in the ZQ subspace during the APHH-CP experiment. (**a**) Different trajectories correspond to different γ angles, increasing from 0 to 2π at steps of π/10. (**b**) The trajectory with γ = 0 (red) is depicted at the bottom, following the representation initiated by Hediger et al.[15] The arrow in grey depicts the trajectory of the Hamiltonian, whose projections on $I_z^{ZQ}$ and $I_x^{ZQ}$ are shown by the dashed arrows.

In a similar fashion, APHH can also be achieved by sweeping across the DQ Hartmann−Hahn matching sideband condition:

$$I_x = I_x^{DQ} + I_x^{ZQ} \xrightarrow{\theta=0 \to \frac{\pi}{2}} I_D^{DQ} + I_x^{ZQ}$$

$$\xrightarrow{\theta=\frac{\pi}{2} \to \pi} -I_x^{DQ} + I_x^{ZQ} = -S_x \quad (28)$$

where

$$I_D^{DQ} \equiv I_z^{DQ} \cos\gamma + I_y^{DQ} \sin\gamma \quad (29)$$

## 2.3 Excitation of MQ coherences from dipolar order

It is interesting to note that the dipolar order $I_D^{ZQ}$ and $I_D^{DQ}$ can be converted to the MQ coherences of the spin system.[25] To demonstrate this point, we transform $I_D^{ZQ}$ from the interaction frame back to the double rotating frame as follows:

$$\rho_D = U_{RF}(\tau)\tilde{\rho}_D U_{RF}^{-1}(\tau) \tag{30}$$

where $\tau$ is the duration of the ADRF process and the unitary operator under the ZQ matching condition is calculated as

$$U_{RF}(\tau) = e^{-i(\omega_{1I}^0 \tau I_x + \omega_{1S}^0 \tau S_x)}$$

$$= e^{-i(\omega_{1I}^0 + \omega_{1S}^0)\tau I_x^{DQ}} e^{-i(\omega_{1I}^0 - \omega_{1S}^0)\tau I_x^{ZQ}}$$

$$= e^{-i(\omega_{1I}^0 + \omega_{1S}^0)\tau I_x^{DQ}} e^{-i\omega_r \tau I_x^{ZQ}} \tag{31}$$

By substituting Equation (25) into Equation (30), we obtain

$$\rho_D = U_{RF}(\tau)\left(I_x^{DQ} + I_D^{ZQ}\right) U_{RF}^{-1}(\tau)$$

$$= I_x^{DQ} + I_z^{ZQ} \cos(\gamma - \omega_r \tau) + I_y^{ZQ} \sin(\gamma - \omega_r \tau)$$

$$= \frac{1}{2}(I_x + S_x) + (I_z S_z + I_y S_y)\cos(\gamma - \omega_r \tau) + (I_y S_z - I_z S_y)\sin(\gamma - \omega_r \tau) \tag{32}$$

When we apply a pair of $(-\pi/2)_y$ pulses on both the $I$ and $S$ channels, ZQ coherences are generated:

$$\rho_D \xrightarrow{\left(-\frac{\pi}{2}\right)I_y} \xrightarrow{\left(-\frac{\pi}{2}\right)S_y} \frac{1}{2}(I_z + S_z) + \rho_{ZQ} \tag{33}$$

where

$$\rho_{ZQ} = (I_x S_x + I_y I_y)\cos(\gamma - \omega_r \tau) + (-I_y S_x + I_x S_y)\sin(\gamma - \omega_r \tau) \tag{34}$$

To excite the DQ coherence, we can choose the DQ matching conditions for the ADRF process so that

$$\rho_D = U_{RF}(\tau)\left(I_D^{DQ} + I_x^{ZQ}\right) U_{RF}^{-1}(\tau)$$

$$= I_z^{DQ} \cos(\gamma - \omega_r \tau) + I_y^{DQ} \sin(\gamma - \omega_r \tau) + I_x^{ZQ}$$

$$= (I_z S_z - I_y S_y)\cos(\gamma - \omega_r \tau) + (I_y S_z + I_z S_y)\sin(\gamma - \omega_r \tau) + \frac{1}{2}(I_x - S_x) \tag{35}$$

When we apply a pair of $(-\pi/2)_y$ pulses on both the $I$ and $S$ channels, DQ coherences are generated:

$$\rho_D \xrightarrow{\left(-\frac{\pi}{2}\right)I_y \left(-\frac{\pi}{2}\right)S_y} \rho_{DQ} + \frac{1}{2}(I_z - S_z) \tag{36}$$

where

$$\rho_{DQ} = (I_x S_x - I_y I_y)\cos(\gamma - \omega_r \tau) + (I_y S_x + I_x S_y)\sin(\gamma - \omega_r \tau) \tag{37}$$

By applying another pair of $(\pi/2)_y$ pulses, the DQ or ZQ coherences can be converted back to the dipolar order.

## 2.4 Direct heteronuclear multiple-quantum spectroscopy

As discussed above, the MQ coherences are excited by the ADRF process followed by a pair of $\pi/2$ pulses. For brevity, we only deal with the chemical shift evolution of the ZQ coherence in the following:

$$\rho_{ZQ}(t_1) = (I_x S_x + I_y I_y)\cos(\gamma - \omega_r \tau + \omega_{ZQ} t_1)$$
$$+ (-I_y S_x + I_x S_y)\sin(\gamma - \omega_r \tau + \omega_{ZQ} t_1) \tag{38}$$

where $\omega_{ZQ} = \omega_S - \omega_I$, as defined in Equation (1) with $p = -1$. After the $t_1$ evolution, the ZQ coherence is converted to the dipolar order by another pair of $\pi/2$ pulses:

$$\frac{1}{2}(I_z + S_z) + \rho_{ZQ} \xrightarrow{\left(\frac{\pi}{2}\right)I_y \left(\frac{\pi}{2}\right)S_y} \rho_D = \frac{1}{2}(I_x + S_x) + (I_z S_z + I_y I_y)\cos(\gamma - \omega_r \tau + \omega_{ZQ} t_1)$$
$$+ (I_y S_z - I_z S_y)\sin(\gamma - \omega_r \tau + \omega_{ZQ} t_1)$$
$$= I_x^{DQ} + I_z^{ZQ}\cos(\gamma - \omega_r \tau + \omega_{ZQ} t_1)$$
$$+ I_y^{ZQ}\sin(\gamma - \omega_r \tau + \omega_{ZQ} t_1) \tag{39}$$

As demonstrated above, we transform the dipolar order again to the interaction frame:

$$\tilde{\rho}_D = U_{RF}^{-1}(\tau + t_1)\rho_D U_{RF}(\tau + t_1)$$
$$= I_x^{DQ} + I_z^{ZQ}\cos(\gamma + \omega_r t_1 + \omega_{ZQ} t_1) + I_y^{ZQ}\sin(\gamma + \omega_r t_1 + \omega_{ZQ} t_1) \tag{40}$$

where

$$U_{RF}(\tau + t_1) = e^{-i(\omega_{1I}^0 + \omega_{1S}^0)(\tau + t_1)I_x^{DQ}} e^{-i\omega_r(\tau + t_1)I_x^{ZQ}} \tag{41}$$

As indicated in Equation (26), only the component parallel to $I_D^{ZQ}$ can be transferred to the Zeeman order. Hence, it is useful to decompose the second term of Equation (40) into the components parallel and perpendicular to the ZQ dipolar order, viz.,

$I_D^{ZQ}$. Accordingly, the $I_D^{ZQ}$ component is calculated as

$$\langle I_z^{ZQ}\cos\gamma(t_1) + I_y^{ZQ}\sin\gamma(t_1)|I_D^{ZQ}\rangle = \langle I_z^{ZQ}\cos\gamma(t_1) + I_y^{ZQ}\sin\gamma(t_1)|I_z^{ZQ}\cos\gamma + I_y^{ZQ}\sin\gamma\rangle$$

$$= \cos(\omega_r t_1 + \omega_{ZQ} t_1) \tag{42}$$

where $\gamma(t_1) \equiv \gamma + \omega_r t_1 + \omega_{ZQ} t_1$ and the $I_{D\perp}^{ZQ}$ component is

$$\langle I_z^{ZQ}\cos\gamma(t_1) + I_y^{ZQ}\sin\gamma(t_1)|I_{D\perp}^{ZQ}\rangle = \langle I_z^{ZQ}\cos\gamma(t_1) + I_y^{ZQ}\sin\gamma(t_1)|I_y^{ZQ}\cos\gamma - I_z^{ZQ}\sin\gamma\rangle$$

$$= \sin(\omega_r t_1 + \omega_{ZQ} t_1) \tag{43}$$

Because the perpendicular component will be dephased by the effective field ($I_e^{ZQ}$) owing to the distribution of the Euler angle $\beta$, it can be ignored in the ARRF process. Thus, the dipolar order described in Equation (40) will be transferred to the Zeeman order through the subsequent ARRF process:

$$\tilde{\rho}_D = I_x^{DQ} + \cos(\omega_r t_1 + \omega_{ZQ} t_1) I_D^{ZQ}$$

$$\xrightarrow{\theta=\frac{\pi}{2}\to\pi} I_x^{DQ} - \cos(\omega_r t_1 + \omega_{ZQ} t_1) I_x^{ZQ} \equiv I_x^{DQ} + \tilde{\rho}_f \tag{44}$$

In the double rotating frame, we have

$$\rho_f = U_{RF}(t)\tilde{\rho}_f U_{RF}^{-1}(t)$$

$$= \cos(\omega_r t_1 + \omega_{ZQ} t_1)\frac{(S_x - I_x)}{2} \tag{45}$$

The $S_x$ magnetization can be directly observed or, in our case, indirectly detected by $^1$H detection after another APHH-CP step. Note that the stationary term $I_x^{DQ}$ can be removed by phase cycling.[26] The phase factor $\omega_r t_1$ describes the phase-time relationship between the prior ADRF and the subsequent ARRF processes.[27] For a rotor-synchronized $t_1$ increment, $\omega_r t_1$ becomes a multiple of $2\pi$. Consequently, we are able to transform the evolved MQ coherences to detectable magnetization $S_x$ and acquire a cosine-modulated signal with frequency $\omega_{ZQ}$ for $p = -1$ in Equation (1). Depending on the choice of the Hartmann–Hahn condition of the APHH-CP step and phase cycling schemes, DQ spectroscopy with $p = +1$ can also be acquired by the same pulse sequence.

    For the quadrature detection in the indirect dimension, we apply the standard hypercomplex method.[28] Briefly, we phase shift the excitation block by $\pi/2$, leading to a corresponding phase shift of the excited MQ coherences. As shown in Equation (43), the detectable term $I_D^{ZQ}$ would then change from cosine modulated to sine modulated. For convenience, the hetMQ experiment conducted by direct excitation of the MQ coherences is denoted as direct-hetMQ in the following discussion. Figure

3a shows the pulse sequence for direct-hetMQ spectroscopy employed in this work.

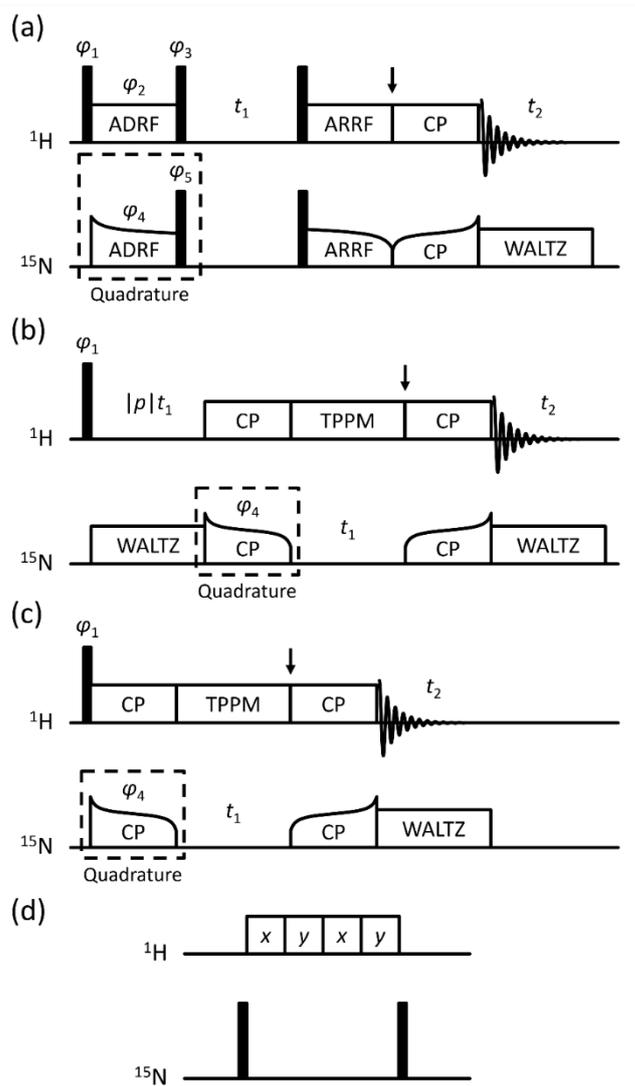

**Figure 3.** Pulse sequences employed in this work. The filled rectangles denote $\pi/2$ pulses. All $^{13}$C decoupling during $t_1$ and $t_2$ evolution periods was omitted for brevity. Suggested phase cycling: for $p > 0$, $\varphi_2 = (0\ 0\ 2\ 2\ 1\ 1\ 3\ 3) \times \pi/2$, $\varphi_4 = (0\ 2\ 0\ 2\ 1\ 3\ 1\ 3) \times \pi/2$, $\varphi_1 = \varphi_2 + \pi/2 = (1\ 1\ 3\ 3\ 2\ 2\ 0\ 0) \times \pi/2$, $\varphi_3 = \varphi_2 - \pi/2 = (3\ 3\ 1\ 1\ 0\ 0\ 2\ 2) \times \pi/2$, $\varphi_5 = \varphi_4 - \pi/2 = (3\ 1\ 3\ 1\ 0\ 2\ 0\ 2) \times \pi/2$, $\varphi_{rec} = (0\ 2\ 2\ 0\ 2\ 0\ 0\ 2) \times \pi/2$; for $p < 0$, $\varphi_2 = (0\ 0\ 2\ 2\ 3\ 3\ 1\ 1) \times \pi/2$ and accordingly $\varphi_1 = \varphi_2 + \pi/2 = (1\ 1\ 3\ 3\ 0\ 0\ 2\ 2) \times \pi/2$, $\varphi_3 = \varphi_2 - \pi/2 = (3\ 3\ 1\ 1\ 2\ 2\ 0\ 0) \times \pi/2$, while $\varphi_2$ and $\varphi_4$ remained the same. Quadrature detection in the indirect dimension using States-TPPI method was implemented by phase shifting the $^{15}$N excitation pulses enclosed by the dashed box by $\pi/2$. (**a**) Direct-hetMQ. (**b**) Concerted-$t_1$ hetMQ spectroscopy. The coherence number $p$ can be non-integer and is independent of the CP modes. (**c**) CP-HSQC spectroscopy. (**d**) Saturation pulses for solvent suppression, if necessary, can be added at the arrow positions in (**a**) to (**c**). The pulses in the $^1$H channel follow the MISSISSIPPI sequence.[44]

## 2.5 Concerted evolution hetMQ spectroscopy

A careful scrutiny of the chemical shift evolution of MQ coherences suggests that the hetMQ correlation spectrum can be obtained without excitation of the MQ coherences. Hereby, we separated the $^1$H and $^{15}$N $t_1$ evolution periods to acquire hetMQ spectra without exciting MQ coherences. The same strategy was well established in the context of coherence transfer echo[29–32] and the split-$t_1$ approach of MQMAS experiments for half-integer quadrupolar nuclei.[33] As shown in Figure 3b, we start with the chemical shift evolution for spin $I$, followed by the APHH-CP step:

$$I_x \xrightarrow{(\omega_I t_1)I_z} I_x \cos(\omega_I t_1) + I_y \sin(\omega_I t_1)$$

$$\xrightarrow{\theta=0\to\pi} S_x \cos(\omega_I t_1) \qquad (46)$$

Note that only the cosine part, which is parallel to $I_x$, can be transferred to $S_x$ by CP. Then, another chemical shift evolution period is set for spin $S$ and the coherence is transferred back to spin $I$ by the second APHH-CP step:

$$S_x \cos(\omega_I t_1) \xrightarrow{(\omega_S t_1)S_z} S_x \cos(\omega_I t_1)\cos(\omega_S t_1) + S_y \cos(\omega_I t_1)\sin(\omega_S t_1)$$

$$\xrightarrow{\theta=\pi\to 0} I_x \cos(\omega_I t_1)\cos(\omega_S t_1) \qquad (47)$$

As a result, the $^1$H detected signal is of double-cosine modulation. By shifting the phases of the pulse block shown in Figure 3, we can create a $\pi/2$ phase shift of the magnetization in Equations (46) and (47). That is, the detected term $I_x$ becomes double sine-modulated signal. These two amplitude-modulated signals can be combined as follows to mimic the chemical shift evolution of MQ coherences:

$$I_x \cos(\omega_I t_1)\cos(\omega_S t_1) + I_x \sin(\omega_I t_1)\sin(\omega_S t_1) = I_x \cos[(\omega_S - \omega_I)t_1] = I_x \cos(\omega_{ZQ} t_1)$$

$$I_x \cos(\omega_I t_1)\cos(\omega_S t_1) - I_x \sin(\omega_I t_1)\sin(\omega_S t_1) = I_x \cos[(\omega_S + \omega_I)t_1] = I_x \cos(\omega_{DQ} t_1) \qquad (48)$$

Note that the selection of ZQ or DQ chemical shift evolution does not depend on the choice of the Hartmann–Hahn matching condition. Similarly, the sine-modulated MQ evolution can be obtained by shifting the phase of the excitation pulses by $\pi/2$. We henceforth refer this pulse sequence to as concerted-$t_1$ hetMQ.

## 2.6 Sheared-HSQC spectroscopy

We have shown that hetMQ correlation spectrum can be obtained without excitation of the MQ coherences. In general, any hetMQ can be obtained by shearing of the corresponding HSQC spectrum.[28] Spectral shearing describes the processing of a 2D spectrum as follows:

$$S(\omega_1, \omega_2) \xrightarrow{\text{shearing}} S(\omega_1 + p\omega_2, \omega_2) \tag{49}$$

Thus, by choosing $p$ equal to +1 or −1, it is trivial to obtain hetMQ spectrum by shearing the HSQC spectrum. As an illustration, a schematic HSQC spectrum was sheared to become the hetDQ and hetZQ spectra (Figure 4). The pulse sequence for the CP-based HSQC experiment is shown in Figure 3c.

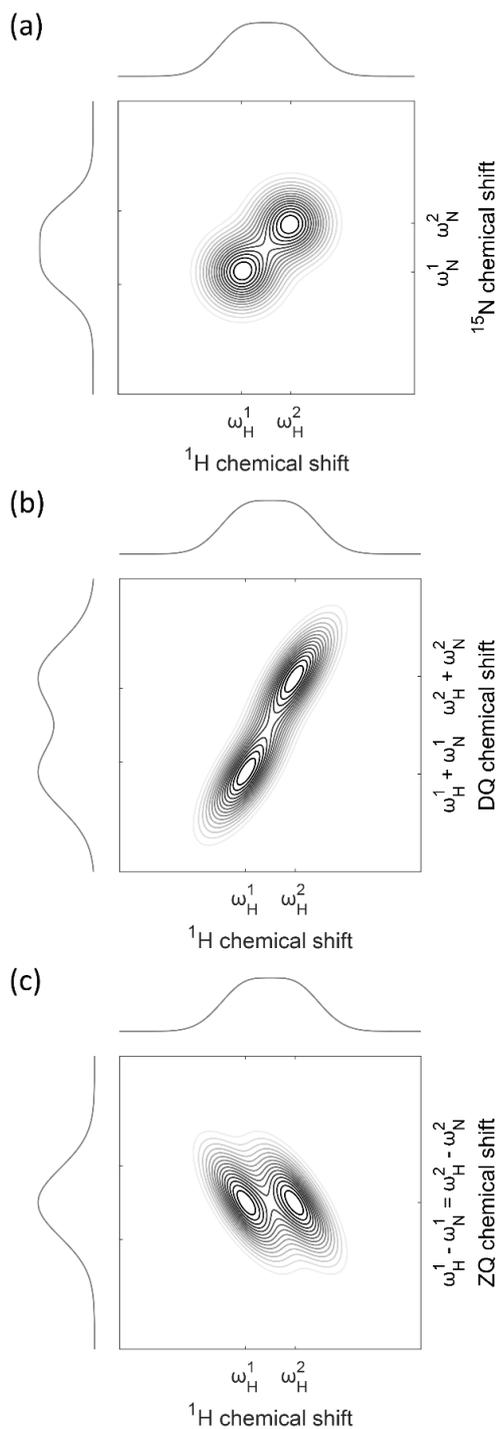

**Figure 4.** Schematic (**a**) HSQC, (**b**) hetDQ, and (**c**) hetZQ spectra. The hetDQ spectrum was obtained by shear transformation of the HSQC spectrum, for which the coherence factor $p$ is +1. The hetZQ spectrum was obtained with $p$ equal to −1. The indirect dimension corresponds to $\omega_H + \omega_N$ and $\omega_H - \omega_N$ for the hetDQ and hetZQ spectra, respectively, where $\omega$ denotes the frequency offset of the peaks.

## 3 RESULTS AND DISCUSSION

The tripeptide of uniformly $^{13}$C and $^{15}$N labeled *N*-formyl-methionyl-leucyl-phenylalanine (fMLF) was used as the model compound in this study. As our reference, the HSQC spectrum of fMLF was acquired at a spinning frequency at 150 kHz (Figure 5). The corresponding DQ and ZQ spectra acquired by three approaches, viz., direct-hetMQ, concerted-$t_1$ hetMQ method, and sheared-HSQC, are shown in Figure 6. As expected, all the signals exhibit the chemical shifts of the MQ coherences in the indirect dimension. Nonetheless, there are considerable differences among the three methods.

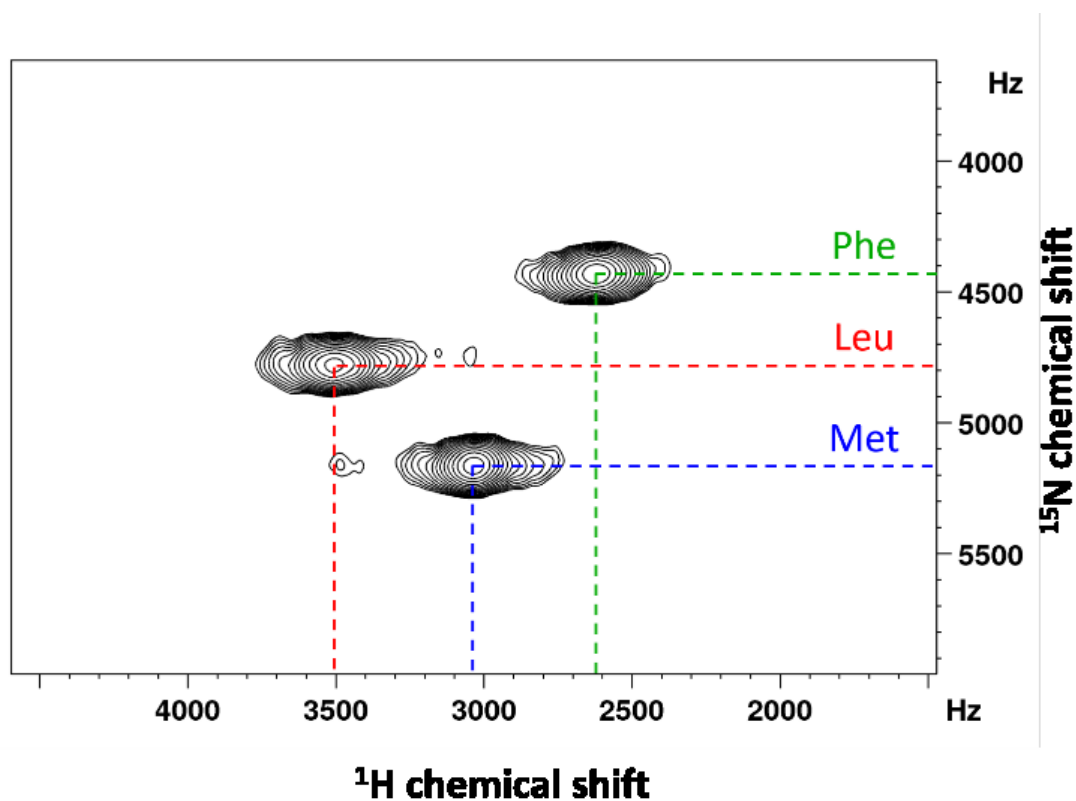

**Figure 5.** $^1$H{$^{15}$N} CP-HSQC spectrum of the uniformly $^{13}$C and $^{15}$N labeled fMLF sample.

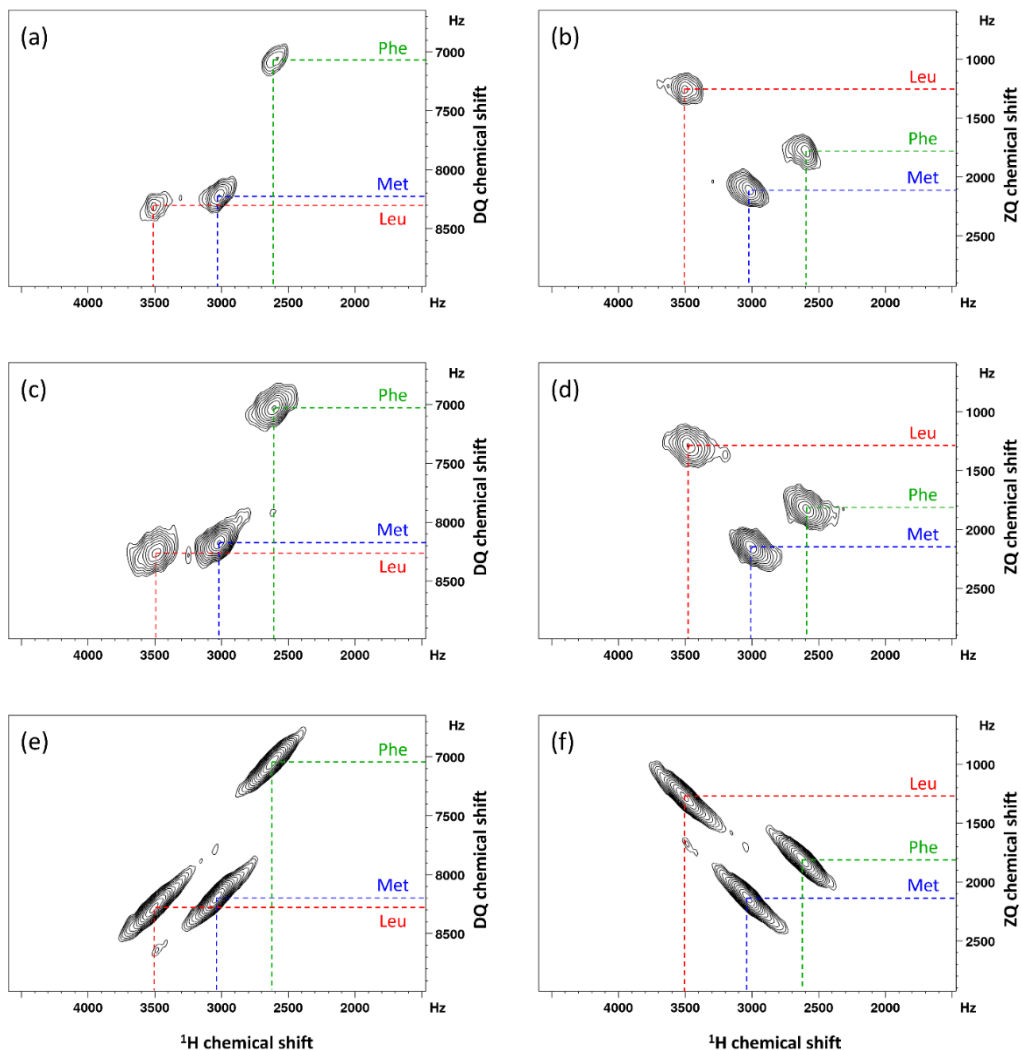

**Figure 6.** 2D $^1$H-$^{15}$N MQ spectra ($p = \pm 1$) of the uniformly $^{13}$C and $^{15}$N labeled fMLF sample acquired by **(a, b)** direct-hetMQ, **(c, d)** concerted-$t_1$ hetMQ pulse sequences, and **(e, f)** sheared HSQC transformation with $p = +1$ and $-1$. The contour levels of all spectra were increased by a factor of 1.198 successively, where the base levels were set to 5 × root-mean-square noise. The $t_1$ evolution of all the spectra was set to 7 ms.

The direct-hetMQ approach is technically the most demanding one. As one can see in Equation (40), the presence of a phase factor of $\omega_r t_1$ will induce a frequency shift of the MQ signal unless the $t_1$ increment is exactly equal to an integer multiple of the rotor period. This phenomenon has also been observed and explained for $^1$H DQ MAS NMR.[34–36] As a verification, we deliberately set a deviation of 500 Hz between the spinning frequency and the inverse of the $t_1$ increment for the hetZQ experiment of fMLF. As shown in Figure 7, the ZQ signals were shifted by 500 Hz in the indirect dimension. By contrast, such an offset between the spinning frequency and the $t_1$ increment did not induce any frequency shift for the signals in the

concerted-$t_1$ hetZQ spectrum. Consequently, the hetMQ experiment using ADRF for direct excitation of the MQ coherences requires the use of a very robust pneumatic control unit. Any fluctuation in the spinning frequency would lead to signal broadening and/or frequency shift. Furthermore, the finite width of the $^1$H and $^{15}$N π/2 pulses may also have deleterious effects on the efficiency of the interconversion between the dipolar order and Zeeman order. Consequently, we observed in Figure 6 that the signal intensity of the direct-hetMQ spectra is relatively low.

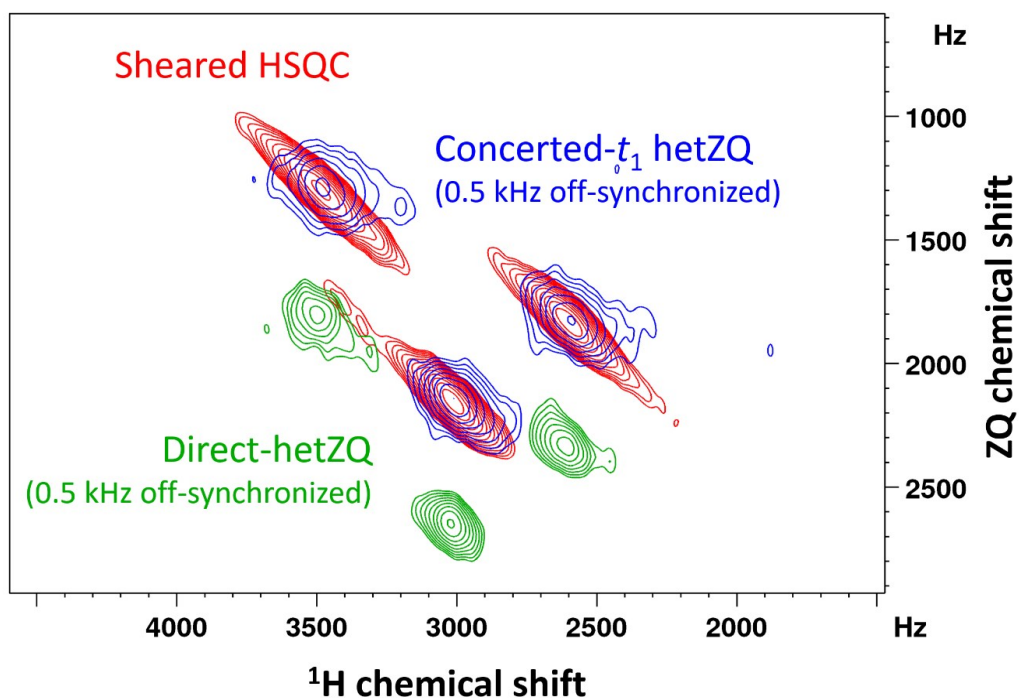

**Figure 7.** Overlay spectrum of hetZQ acquired under a 0.5 kHz deviation between the inverse of $t_1$ increment and the spinning frequency for the uniformly $^{13}$C and $^{15}$N labeled fMLF sample. The $t_1$ increment was set to 64 × (150 kHz)$^{-1}$ = 427 μs, and the spinning frequency was set to 149.5 kHz. The direct-hetZQ spectrum (green) was shifted by 0.5 kHz, whereas the concerted-$t_1$ hetZQ spectrum (blue) remained unshifted. The sheared HSQC spectrum (red) was shown as a reference spectrum.

The technical challenges of direct-hetMQ are alleviated in concerted-$t_1$ hetMQ experiments. Because no MQ coherences are excited, the choice of the Hartmann–Hahn matching condition becomes immaterial. At a spinning frequency of 150 kHz, we found that the ZQ matching condition $\omega_I^0 - \omega_S^0 = \omega_r$ had a slightly better CP efficiency than the DQ matching condition. The two concerted-$t_1$ evolution periods allow the application of heteronuclear decoupling, which may lengthen the lifetime of the coherences. Hence, it is not surprising that the signal intensity of concerted-$t_1$ hetMQ is considerably better than that of direct-hetMQ. Furthermore, an interesting merit of the concerted-$t_1$ hetMQ spectroscopy is that the coherence factor $p$ can go beyond $\pm 1$, simply by adjusting the ratio between the $t_1$ increments of the two evolution periods.

For a 2D spectrum, the spectral resolution between two signals can be defined by Equation (2) with respect to the tilted axis connecting the two signals under consideration (Figure S1). As shown in the CP-HSQC spectrum, the three axes connecting the signal pairs of Met-Leu, Phe-Met, and Leu-Phe have different slopes. Hence, a particular $p$ value can at best provide the optimal resolution for only one of the three signal pairs. In other words, the spectral resolution in the indirect dimension, as a function of the coherence number $p$ in Equation (1), cannot be optimized simultaneously for all signal pairs. Figure 8 illustrates the results of concerted-$t_1$ hetMQ spectra for $p$ ranging from +4 to −4. As expected, the optimal resolution in the indirect dimension was achieved in the concerted-$t_1$ hetDQ and hetZQ spectra for the signal pairs of Leu-Phe and Met-Leu, respectively. Similar results were obtained by shearing the CP-HSQC spectra accordingly (Figure S2).

In the presence of inhomogeneous broadening,[37] which is commonly found in samples with substantial structural distribution, the resolution of the projection onto the indirect dimension, viz., $F_1$ dimension, of the hetMQ spectrum may be superior to that of the corresponding HSQC spectrum. In the following, we demonstrated that it is very flexible to obtain various pseudo hetMQ spectra by shearing the HSQC spectrum. Previously, we reported the $^1$H{$^{13}$C} HSQC spectrum acquired for spruce.[9] As reproduced in Figure 9a, the experimental CP-HSQC $^1$H-$^{13}$C signal pairs do not run parallel to spectral axes. The correlation ridges revealed that the line widths of the $^1$H NMR peaks were dominated by the variation in chemical shifts but not relaxation effects. A visually similar tilting of correlated peaks has been observed in the context of the correlation between high-order and single-quantum coherences[38–40] and anisotropic bulk magnetic susceptibility generated by aromatic rings.[41] Although there is no compelling physical reason why composite signal should follow a straight line in $F_1$-$F_2$ plane, we use this approximation for resolution enhancement and perform shear transformation at various tilt angles for each site (Figure 9b). Transformed plots allow a selection of the narrowest projection for any tilted line and improve overall resolution, while vertical axis gives an angle of shearing. It is remarkable that the optimum tilting of each correlation peak, which carries individual structural information associated with the chemical moieties, was different from the others.

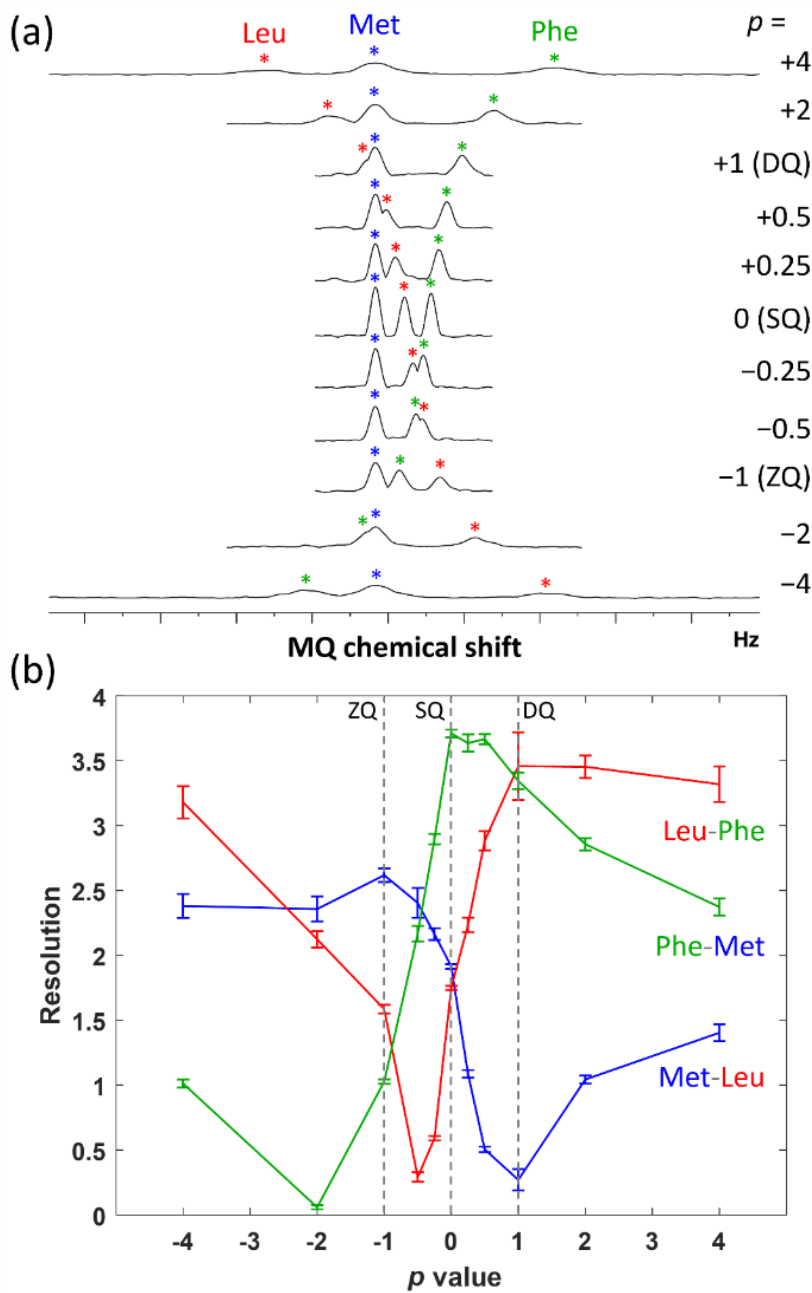

**Figure 8.** Spectral resolution on the indirect dimension of hetMQ spectra of the uniformly $^{13}$C and $^{15}$N labeled fMLF sample. (**a**) Skyline projections of the indirect dimension of the concerted-$t_1$ hetMQ spectroscopy experiments. For comparison, all projections are aligned with respect to the methionine (Met) signal of the HSQC spectrum ($p = 0$). The spectral windows for $p = \pm 2$ and $\pm 4$ were expanded to accommodate the whole spectra, while the $t_1$ acquisition time remained unchanged. The peak broadening was attributed to the signal decay during the $^1$H $t_1$ period. (**b**) The resolution of signal pairs, defined by the spectral distances divided by the line widths, as a function of coherence factor $p$. The error bars indicate the 95% confidence interval calculated from the fitting results. The dashed lines indicate the $p$ values corresponding to the hetZQ, HSQC, and hetDQ spectra.

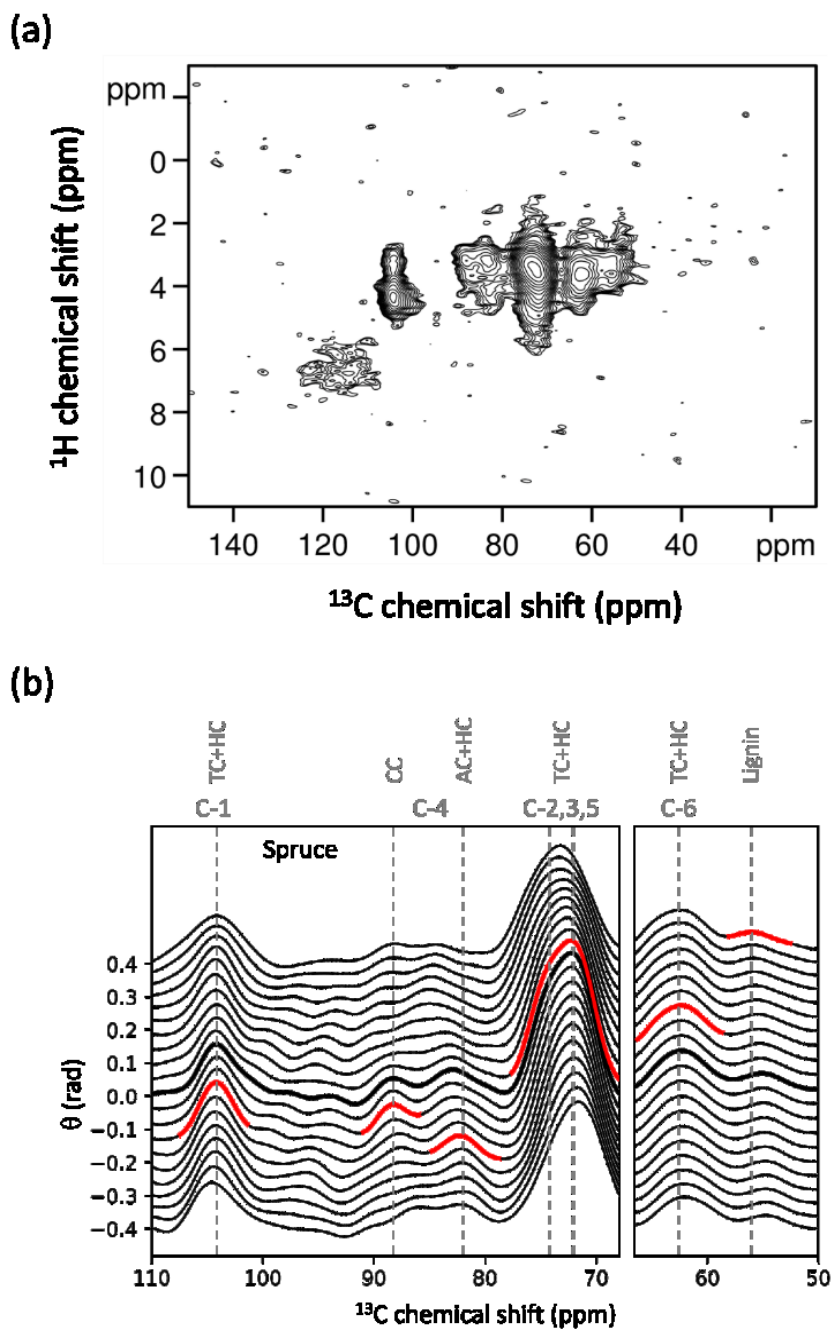

**Figure 9.** (a) $^1\text{H}\{^{13}\text{C}\}$ CP-HSQC spectrum of spruce of natural abundance acquired at 18.8 T under a MAS frequency of 140 kHz (Reproduced with permission from Figure S2 of reference 9). (**b**) $^{13}\text{C}$ projections of the $^1\text{H}\{^{13}\text{C}\}$ CP-HSQC spectra after the Radon transformation (shear + sum projection) at various tilt angles ($\theta$).[45] The projections in thickened black lines are the unsheared projections. The peaks with optimal resolution were highlighted in red. The dashed lines are eye guides for the optimal peak positions. The spectral region was split into two because of their slightly different spans in the sum projections. HC: hemicellulose. AC/CC/TC: amorphous/crystalline/total cellulose. C1 to C6 refer to the carbons of the cellulose framework.

We note in passing that the correspondence between MQ-SQ correlation spectrum and the sheared SQ-SQ correlation spectrum may also apply to homonuclear systems. As an illustration, a homonuclear SQ-SQ correlation spectrum can be transformed to a DQ-SQ correlation spectrum by shearing (Figure S3). However, it should be cautioned that a homonuclear DQ-SQ correlation spectrum could be different from a sheared SQ-SQ correlation spectrum because the excitation of the auto-correlation peaks, which depends on the spin geometry and intermolecular distances, may not be always feasible.

## 4   CONCLUSION

Heteronuclear MQ-SQ correlation or hetMQ spectra are equivalent to various shear transformations of the HSQC spectrum in the $F_1$ dimension. Thus, hetMQ spectra do not provide more spectral information than HSQC spectra. Only if a signal is resolved in a HSQC spectrum, it can possibly be resolved in the hetMQ spectra. Although the direct-hetMQ and concerted-$t_1$ hetMQ methods may appear superfluous because hetMQ spectra can be obtained by shearing of the HSQC spectrum, for the kind of 2D correlation spectroscopy between, say, $^1H$-$^{15}N$ MQ and $^{13}C$ SQ coherences, it is more straightforward and advantageous to acquire the MQ chemical shift evolution using the method of concerted-$t_1$ hetMQ.

## 5   EXPERIMENTAL

The uniformly $^{13}C$- and $^{15}N$-labeled fMLF was purchased from CortecNet (Les Ulis, France). All spectra were acquired at 9.4 T ($^{13}C$ and $^1H$ Larmor frequencies of 100.6 and 400.1 MHz, respectively) on a Bruker Avance III spectrometer equipped with a NMRI/Darklands built 0.51-mm probe. The MAS rate was 150 kHz unless stated otherwise. The rotor volume of approximately 200 nL was fully packed with fine powder for each measurement. For all the experiments, the duration of $^1H$ and $^{15}N$ 90° pulses were set to 1 and 3 μs, respectively. For the CP-HSQC and concerted-$t_1$ hetMQ methods, the initial CP contact time was set to 2 ms and the second contact time was 600 μs. For the direct-hetMQ method, the ADRF and ARRF contact times were 1 ms. The following CP contact time was 600 μs. During the contact time, the $^1H$ nutation frequency was set to 100 kHz for the direct-hetDQ spectrum and 200 kHz for all other spectra, while that of $^{15}N$ was tangentially swept from 40 to 60 kHz as follows:

$$\omega_1(t) = A \times \frac{\tan\left(\frac{2t-\tau}{\tau} \times \theta\right)}{\tan\theta}$$

where $\tau$ is the contact time, $A$ and $\theta$ were set to 10 kHz and 80°. $^1$H, $^{13}$C and $^{15}$N decoupling were applied using TPPM-15 at a $B_1$ field of 37.5 kHz, WALTZ-16 of 10 kHz and WALTZ-16 of 5 kHz, respectively.[42,43] A total of 16 increments were acquired at steps of 427 μs (64 rotor periods), corresponding to spectral width of 2343.75 Hz. The recycle delay was set to 1 s. The number of transients accumulated for each $t_1$ increment was 32. Quadrature detection in the indirect dimension was achieved by the States-TPPI method.[28]

All the spectra were processed with an exponential window function (line broadening 100 Hz) in the $F_2$ dimension and a squared sine bell shifted by π/2 in the $F_1$ dimension. The shear transformation was realized in TopSpin 4.0.8 with the built-in command *ptilt1* to tilt the spectra along columns. The spectral deconvolutions of the MQ skyline projections were carried out with pure Gaussian line shapes and a constant baseline.

# 6 ACKNOWLEDGEMENT

This work was financially supported by the Ministry of Science and Technology (MOST) through the grants of 107-2113-M-002-004 and 108-2628-M-002-012-MY3. The NMR measurements were carried out at the Instrumentation Center of National Taiwan University. ECYY thanks the support of MOST for an undergraduate research scholarship. MLO and AS acknowledge TUT Rector Fond for financial support.

# 7 REFERENCES

1. S. P. Brown. *Solid State Nucl. Magn. Reson.* **2012**, *41*, 1–27.

2. R. Zhang, K. H. Mroue, A. Ramamoorthy. *Acc. Chem. Res.* **2017**, *50* (4), 1105–1113.

3. Y. Ishii, R. Tycko. *J. Magn. Reson.* **2000**, *142*, 199–204.

4. Y. Ishii, J. P. Yesinowski, R. Tycko. *J. Am. Chem. Soc.* **2001**, *123*, 2921–2922.

5. A. Samoson. *J. Magn. Reson.* **2019**, *306*, 167–172.

6. F. M. Paruzzo, B. J. Walder, L. Emsley. *J. Magn. Reson.* **2019**, *305*, 131–137.

7. P. Moutzouri, F. M. Paruzzo, B. Simões de Almeida, G. Stevanato, L. Emsley. *Angew. Chem. Int. Ed.* **2020**, *59* (15), 6235–6238.

8. Y.-L. Lin, Y.-S. Cheng, C.-I. Ho, Z.-H. Guo, S.-J. Huang, M.-L. Org, A. Oss, A. Samoson, J. C. C. Chan. *Chem. Commun.* **2018**, *54* (74), 10459–10462.

9. E. C.-Y. Yuan, S.-J. Huang, H.-C. Huang, J. Sinkkonen, A. Oss, M.-L. Org, A. Samoson, H.-C. Tai, J. C. C. Chan. *Chem. Commun.* **2021**, *57* (34), 4110–4113.

10. W. Sommer, J. Gottwald, D. Demco, H. Spiess. *J. Magn. Reson. A* **1995**, *113* (1), 131–134.

11. M. Hong, J. D. Gross, R. G. Griffin. *J. Phys. Chem. B* **1997**, *101* (30), 5869–5874.

12. K. Saalwachter, R. Graf, D. E. Demco, H. W. Spiess. *J. Magn. Reson.* **1999**, *139* (2), 287–301.

13. K. Saalwächter, R. Graf, H. W. Spiess. *J. Magn. Reson.* **2001**, *148* (2), 398–418.

14. K. Saalwächter, H. W. Spiess. *J. Chem. Phys.* **2001**, *114* (13), 5707–5728.

15. S. Hediger, B. H. Meier, N. D. Kurur, G. Bodenhausen, R. R. Ernst. *Chem. Phys. Lett.* **1994**, *223* (4), 283–288.

16. S. Hediger, B. H. Meier, R. R. Ernst. *Chem. Phys. Lett.* **1995**, *240* (5), 449–456.

17. M. Bjerring, J. T. Rasmussen, R. Schultz Krogshave, N. Chr. Nielsen. *J. Chem. Phys.* **2003**, *119* (17), 8916–8926.

18. M. Bjerring, N. Chr. Nielsen. *Chem. Phys. Lett.* **2003**, *382* (5), 671–678.

19. A. Wokaun, R. R. Ernst. *J. Chem. Phys.* **1977**, *67* (4), 1752–1758.

20. S. Vega. *J. Chem. Phys.* **1978**, *68* (12), 5518–5527.

21. N. Khaneja, C. Kehlet, S. J. Glaser, N. Chr. Nielsen. *J. Chem. Phys.* **2006**, *124* (11), 114503.

22. M. H. Levitt. In *Encyclopedia in Nuclear Magnetic Resonance*; Edited by D. M. Grant and R. K. Harris; Wiley: Chichester, 2002; Vol. 9, pp 165–196.

23. A. Pines, M. G. Gibby, J. S. Waugh. *J. Chem. Phys.* **1973**, *59* (2), 569–590.

24. J. Jeener, P. Broekaert. *Phys. Rev.* **1967**, *157* (2), 232–240.

25. J. D. van Beek, A. Hemmi, M. Ernst, B. H. Meier. *J. Chem. Phys.* **2011**, *135* (15), 154507.

26. D. Marion, M. Ikura, R. Tschudin, A. Bax. *J. Magn. Reson.* **1989**, *85* (2), 393–399.

27. Levitt, M. H. Symmetry-Based Pulse Sequence in Magic-Angle Spinning Solid-State NMR. In *Encyclopedia in Nuclear Magnetic Resonance*; Edited by D. M. Grant and R. K. Harris; Wiley: Chichester, 2002; Vol. 9, pp 165–196.

28. R. R. Ernst, G. Bodenhausen, A. Wokaun. *Principles of Nuclear Magnetic Resonance in One and Two Dimensions*; Clarendon Press: Oxford, 1990.

29. D. P. Weitekamp, J. R. Garbow, J. B. Murdoch, A. Pines. *J. Am. Chem. Soc.* **1981**, *103* (12), 3578–3579.

30. P. H. Bolton, G. Bodenhausen. *J. Magn. Reson.* **1982**, *46* (2), 306–318.

31. S. L. Duce, L. D. Hall, T. J. Norwood. *J. Magn. Reson.* **1990**, *89* (2), 273–286.

32. S. L. Duce, L. D. Hall, T. J. Norwood. *Chem. Commun.* **1990**, No. 2, 146–148.

33. S. P. Brown, S. Wimperis. *J. Magn. Reson.* **1997**, *124* (1), 279–285.

34. H. Geen, J. J. Titman, J. Gottwald, H. W. Spiess. *J. Magn. Reson. A* **1995**, *114* (2), 264–267.

35. U. Friedrich, I. Schnell, D. E. Demco, H. W. Spiess. *Chem. Phys. Lett.* **1998**, *285* (1–2), 49–58.

36. S. P. Brown, A. Lesage, B. Elena, L. Emsley. *J. Am. Chem. Soc.* **2004**, *126* (41), 13230–13231.

37. M. Maricq, J. Waugh. *J. Chem. Phys.* **1979**, *70* (7), 3300–3316.

38. D. Sakellariou, S. P. Brown, A. Lesage, S. Hediger, M. Bardet, C. A. Meriles, A. Pines, L. Emsley. *J. Am. Chem. Soc.* **2003**, *125* (14), 4376–4380.

39. S. Cadars, A. Lesage, L. Emsley. *J. Am. Chem. Soc.* **2005**, *127* (12), 4466–4476.

40. T. Kobayashi, K. Mao, P. Paluch, A. Nowak-Król, J. Sniechowska, Y. Nishiyama, D. T. Gryko, M. J. Potrzebowski, M. Pruski. *Angew. Chem. Int. Ed.* **2013**, *52* (52), 14108–14111.

41. M. P. Hanrahan, A. Venkatesh, S. L. Carnahan, J. L. Calahan, J. W. Lubach, E. J. Munson, A. J. Rossini. *Phys. Chem. Chem. Phys.* **2017**, *19* (41), 28153–28162.

42. A. E. Bennett, C. M. Rienstra, M. Auger, K. V. Lakshmi, R. G. Griffin. *J. Chem. Phys.* **1995**, *103*, 6951–6958.

43. A. J. Shaka, J. Keeler, R. Freeman. *J. Magn. Reson.* **1983**, *53* (2), 313–340.

44. D. H. Zhou, C. M. Rienstra. *J. Magn. Reson.* **2008**, *192* (1), 167–172.

45. A. G. Rann, A. I. Katsevich. *The Radon Transform and Local Tomography*; CRC Press, 1996.

# Supporting Information

**Solid-state heteronuclear multiple-quantum spectroscopy under a magic-angle spinning frequency of 150 kHz**

Eric Chung-Yueh Yuan,[a] Po-Wen Chen,[a] Shing-Jong Huang,[b] Mai-Liis Org,[c] Ago Samoson,[c,*] and Jerry Chun Chung Chan[a,*]

[a] Department of Chemistry, National Taiwan University, Taipei 10617, Taiwan, Republic of China

[b] Instrumentation Center, National Taiwan University, Taipei 10617, Taiwan, Republic of China

[c] Tallinn University of Technology, Estonia

---

[*] Corresponding authors: ago.samoson@ttu.ee, chanjcc@ntu.edu.tw

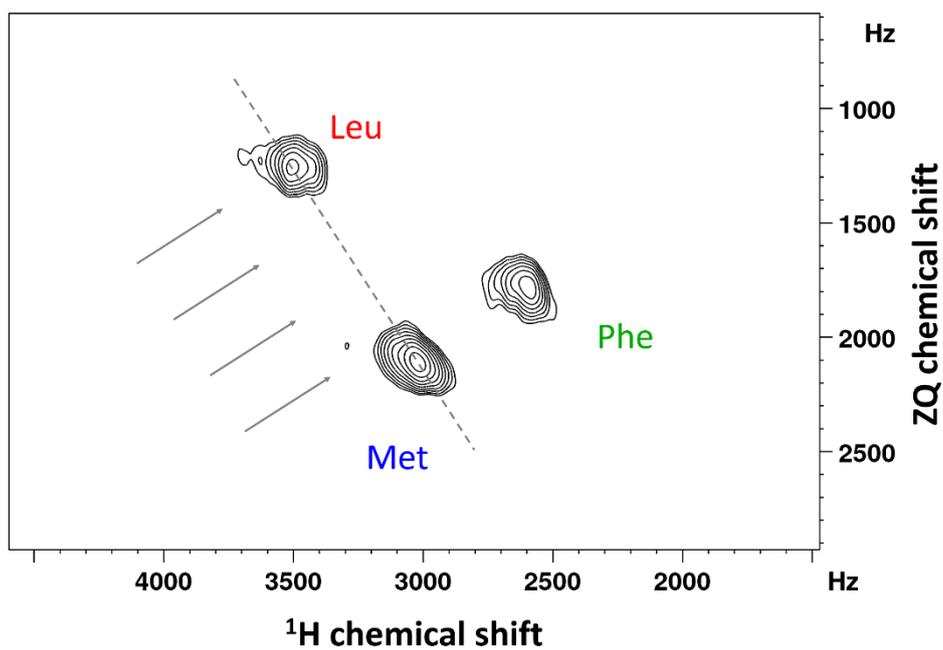

**Figure S1.** Illustration of the optimal projection direction for the signal pair of Leu and Met. The directions are indicated by the arrows. The dashed line depicts the tilted axis connecting the signals of Leu and Met. The arrows are perpendicular to the tilted axis. Along the optimal direction, the projected signals have the highest spectral resolution.

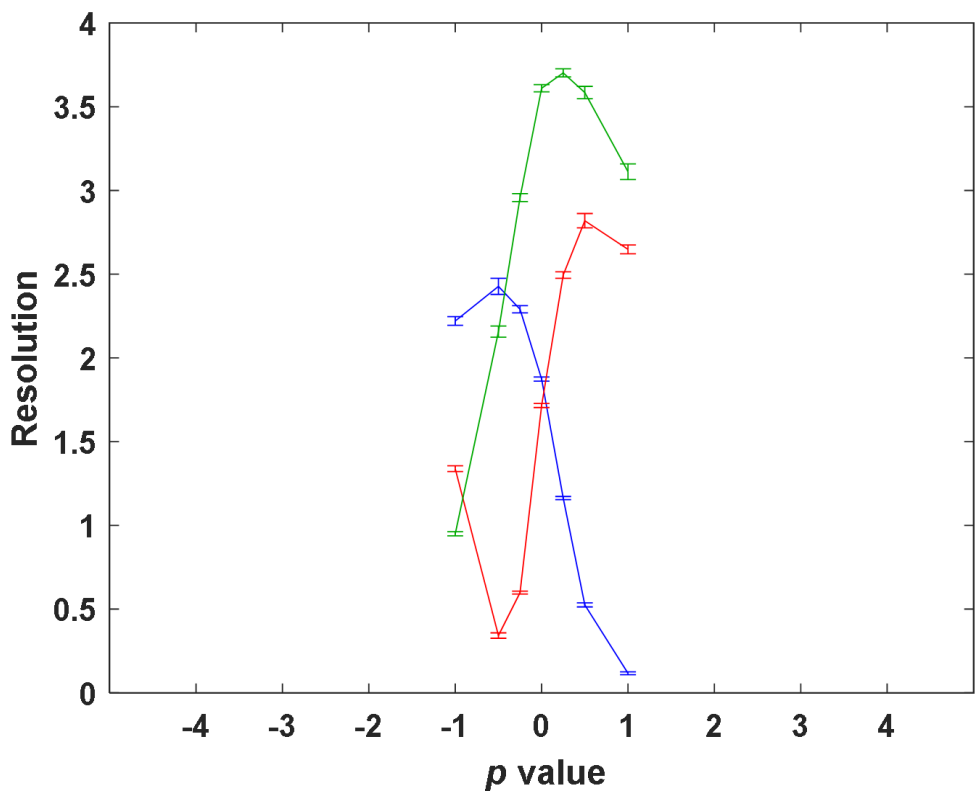

**Figure S2.** Resolution of the signal pairs on the indirect dimension of the sheared HSQC spectrum of the uniformly $^{13}$C and $^{15}$N labeled fMLF sample. The color code is identical to that shown in Figure 8 of the main text. The spectral window in F$_1$ was insufficient for the shearing with other *p* values.

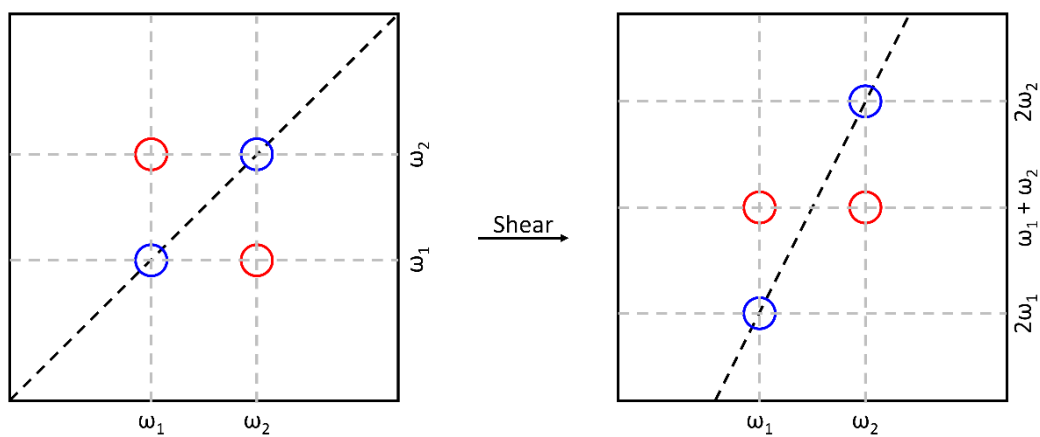

**Figure S3.** Homonuclear SQ-SQ correlation spectrum is transformed to DQ-SQ correlation spectrum by shearing with *p* = +1.